\documentclass[onecolumn,preprintnumbers,amsmath,amssymb]{revtex4}
\usepackage{epsfig,graphicx}
\usepackage{dcolumn}
\usepackage{bm}
\begin{document}
\draft
\title{Electronic properties of NiCl$_2$ tubular nanostructures}

\vspace{1cm}
\author{V.V. Ivanovskaya*, A.N. Enyashin, N.I. Medvedeva and A.L. Ivanovskii\footnote[1]{For correspondence: viktoria@ihim.uran.ru}}
\address{Institute of Solid State Chemistry,\\ Ural Branch of the Russian Academy of
Sciences,\\Ekaterinburg, 620219, Russia}

\begin{abstract}

Atomic models of {\it zigzag} ({\it n}, 0)- and {\it armchair}
({\it n},{\it n})-like NiCl$_2$ nanotubes (n = 4 - 29) formed by
rolling (100) single layers of the bulk NiCl$_2$ which
crystallizes in the CdCl$_2$-type structure, are constructed and
their electronic properties and bond indices are investigated
using the tight-binding band theory. The calculations performed
show that all the considered nanotubes in non-magnetic state are
uniformly metallic-like. The density of states at the Fermi level
contains a considerable contribution from Ni3d states, its value
depends on the atomic configuration and diameter of the tubes.
The Ni-Cl covalent bonds were found in NiCl$_2$ tubes, whereas
Ni-Ni and Cl-Cl covalent interactions are almost absent. According
our estimations, the {\it zigzag} ({\it n}, 0)-like NiCl$_2$
nanotubes are more energetically favorable.

\end{abstract}
\maketitle


Quasi-one-dimensional (1{\it D}) inorganic nanotubes (NTs) attract
much interest due to wide prospects of technological application
of materials produced on their basis
\cite{dresselhaus,saito,moriarty,knupfer,ajayan,tenne,ivanovskii,pokropivnyi}.
Besides carbon (see \cite{dresselhaus,saito,moriarty}), the first
inorganic compounds, which were the basis for nanotube
production, were layered hexagonal boron nitride
\cite{copra,louseau,rubio} and d-metal dichalcogenides MX$_2$ (M =
Mo, W, Ta; X = S, Se)
\cite{tenne,nath,rosentsveig,tenne2,tenne3}, having structures
analogous to that of graphite. Recently, NTs of other sulfides
were successfully synthesized: InS \cite{hollingsworth}, ZnS
\cite{wang}, Bi$_2$S$_3$ \cite{ye}, ReS$_2$ \cite{bronson},
W$_x$Mo$_y$C$_z$S$_2$ \cite{hsu}, ZrS$_2$, HfS$_2$
\cite{nath2,nath3}.

Along with the above-mentioned NTs, a large number of other
inorganic nanotubes have been produced or predicted. Among them
there are new 1{\it D} tubular structures of metals (Na, Cu
\cite{opitz}, Ni \cite{bao}, Ag, Au \cite{lay,ahn}, Bi
\cite{li,su}), boron \cite{boustani}, and silicon \cite{zhang}.
Some efforts are reported in synthesis and modeling of nanotubes
based on inorganic compounds with planar or pseudoplanar
structures: BC$_3$, BC$_2$N, CN$_x$ (see reviews
\cite{tenne,ivanovskii,pokropivnyi}), metal diborides MB$_2$ (M =
Mg, Al, Sc, Ti, Zr) \cite{chernozatonskii,quandt,ivanovskaya},
AlB$_2$-like ternary silicides Sr(Ga$_x$Si$_1$$_-$$_x$)$_2$,
Ca(Al$_x$Si$_1$$_-$$_x$)$_2$ \cite{shein} etc. A large number of
1{\it D} nanomaterials (nanowires, nanorods, nanotubes) based on
metal oxides such as TiO$_2$
\cite{imai,shimizu,kasuga,seo,miao,yao}, V$_2$O$_5$
\cite{satishkumar,krumeich,muhr,niederberger,dobley,pan,chandrappa},
Co$_3$O$_4$ \cite{shi,shi2}, MnO$_2$, WO$_3$ \cite{lakshmi},
MoO$_3$, ReO$_2$ \cite{satishkumar2} and also on ZnO \cite{hu},
SiO$_x$ \cite{zhang2,wang2} etc. have been obtained. Important
progress was achieved in producing nanotubes of semiconducting
materials: SiGe, InAs/GaAs, InGaAs/GaAs, SiGe/Si InGeAs/GaAs
\cite{schmidt,prinz,deneke}.

Recently, the synthesis of closed-cage nanoclusters and nanotubes
NiCl$_2$ which is the first reprsentative of layered
metal-halogen compounds, was reported \cite{hacohen,hacohen2}.
High-resolution transmission electron microscopy analysis (HRTEM)
and electron diffraction measurements showed that NiCl$_2$ layers
forming the walls of those multi-walled NTs were completely
crystalline. Taking into account that bulk NiCl$_2$ is of ferro-
and weak antiferromagnetic Ni-Ni coupling type in and across the
layers, respectively, it is suggested \cite{hacohen} that unique
magnetic properties may be expected for NiCl$_2$ NTs.

In this communication, the atomic simulation of the nanotubes
constructed from NiCl$_2$ single layers is performed and their
electronic band structure and bond indices are calculated and
analyzed as a function of the tubes diameters (D) in {\it
armchair}- and {\it zigzag}-like forms.

NiCl$_2$ crystallizes in the CdCl$_2$ structural type (space group
{\it P3m}) with the lattice parameters {\it a} = 0.3483, {\it c}
=1.7400 nm. The bulk NiCl$_2$ consists of layers of a Ni sheet
sandwiched between two chlorine sheets (Cl-Ni-Cl). Within the
layers there exist ionic-covalent Ni-Cl bonds; the neighboring
(Cl-Ni-Cl) layers interact with each other due to weak Van der
Waals forces. The atomic models of tubular NiCl$_2$ structures
were constructed analogously to the nanotubes of layered Mo, W,
Nb dichalcogenides
\cite{seifert,seifert2,seifert3,seifert4,seifert5} by mapping a
"triple" (Cl-Ni-Cl) layer onto the tube surface. Such
"single-walled" tubes consist of three coaxial Cl-Ni-Cl cylinders
built up of wrapped hexagonal Ni, Cl monoatomic sheets, Fig. 1.
As for single-walled carbon NTs, see
\cite{dresselhaus,saito,moriarty,knupfer,ajayan,tenne,ivanovskii,pokropivnyi},
three groups of NiCl$_2$ NTs can be classified as non-chiral ({\it
armchair} ({\it n},{\it n})-, {\it zigzag} ({\it n},0)-like) and
chiral ({\it n},{\it m})-like nanotubes. The majority of the
synthesized \cite{hacohen2} NiCl$_2$ NTs were non-chiral.

We calculated the electronic structure of non-chiral ({\it n},0)
and ({\it n},{\it n}) NiCl$_2$ NTs as a function of {\it n} in
ranges from (8,0) to (29,0) and from (4,4) to (29,29), which
correspond to the intervals of the diameters of "outer" chlorine
cylinders (D$_o$$_u$$_t$) 2.272 - 7.096 and 1.984 - 11.902 nm,
respectively, Table 1.

The tight binding band structure method within the extended
Huckel theory (EHT) approximation \cite{hoffmann} was employed.
Besides the electronic spectra, the densities of states (DOS),
crystal orbital overlap populations (COOPs), and total band
energies of the nanotubes ({\it E}$_t$$_o$$_t$) were obtained.

Figure 2 shows the calculated values of {\it E}$_t$$_o$$_t$ (per
NiCl$_2$ unit) versus NT diameters. The {\it E}$_t$$_o$$_t$
dependence follows a ~ 1/D$^2$ behavior indicative of a decrease
in NTs stability with decreasing D. Analogous dependences of
strain energy (the difference between the energies of the plane
atomic layer and the corresponding NT characterizes the chemical
stability of tubular structures) are known for carbon and the
majority of non-carbon NTs
\cite{dresselhaus,saito,moriarty,knupfer,ajayan,tenne,ivanovskii,pokropivnyi}.
It is worth noting that according to our results for all the
considered tubes, {\it zigzag}-like configurations of the NiCl$_2$
nanotubes are more stable.

Within non-magnetic approach, all NTs are metallic-like. The
electronic DOSs for the (15, 15) and (15, 0) NiCl$_2$ nanotubes
are showed in Fig. 3. The pronounced very sharp DOS peak near the
Fermi level  consists mainly of Ni3d states. Hence the magnetic
instability can occur and it may be assumed that the near-Fermi
Ni3d states will play a large role in the magnetic properties of
NiCl$_2$ NTs. However this aspect remains to be considered.

It is significant that the shape of the electronic spectrum is
essentially modified by the tube geometry, see Fig. 3. For
example, for {\it armchair} (15, 15) NTs, the near-Fermi bands
are separated from the filled valence bands by a forbidden gap,
which is absent for {\it zigzag} (15, 0) NTs.

To analyze the covalence of intra-atomic interactions, the COOPs
of paired bonds were examined. It was established that Ni-Cl
covalent ineractions are the strongest bonds (Table 2), whereas
Ni-Ni and Cl-Cl covalent-type interactions are absent (COOPs less
0). There is also sharp anisotropy of separate Ni-Cl bonds
depending both on their orientation relative to the NT axis and
the coordination of chlorine atoms belonging to "inner" or
"outer" cylinders, as well as on the tube geometry. It is seen
that the ratio of COOPs values for Ni-Cl$_i$$_n$ and
Ni-Cl$_o$$_u$$_t$ bonds differs drastically for {\it armchair}
({\it n},{\it n}) - and {\it zigzag} ({\it n},0)-like tubes. In
particular, for {\it zigzag}-like tubes the COOPs values of
Ni-Cl$_i$$_n$ are much greater than those for Ni-Cl$_o$$_u$$_t$.
This suggests higher reactivity of the "outer" chlorine atoms and
a possibility of their rearrangement. The above features of
interatomic bonds depending on the tube geometry may become an
important factor in developing microscopic models of NiCl$_2$
tubes growth both in {\it zigzag}- and {\it armchair}-like (and
also chiral \cite{hacohen2}) forms.

In conclusion we have constructed atomic models of single-walled
NiCl$_2$ nanotubes in {\it zigzag}- and {\it armchair}-like forms
and investigated their electronic properties and bond indices by
the tight-binding band method. We show that both {\it zigzag}- and
{\it armchair}-like nanotubes in non-magnetic state are
metallic-like, with a very sharp DOS peak near the Fermi level
predominantly of Ni3d character. This points to magnetic
instability and a possibility of magnetic phenomena in NiCl$_2$
NTs. {\it Zigzag}-like NiCl$_2$ nanotubes were found to be more
stable. It was established that Ni-Cl covalent bonds are the
strongest interactions in nickel dichloride NTs.

\newpage

\begin{table}
\begin{center}
\caption{Radii (R, nm)* for NiCl$_2$ nanotubes.}
\begin{center}
\begin{tabular}{|c|c|c|c|c|c|c|c|}
\hline Nanotubes& R$^i$$^n$(Cl)& R$^o$$^u$$^t$(Cl)& R(Ni)&Nanotubes& R$^i$$^n$(Cl)& R$^o$$^u$$^t$(Cl)& R(Ni)\\
\hline (8,0)&0,535&1,136&0,904&(6,6)&0,802&1,405&1,174\\
\hline (9,0)&0,657&1,254&1,017&(7,7)&1,011&1,607&1,370\\
\hline (10,0)&0,778&1,371&1,130&(8,8)&1,216&1,809&1,566\\
\hline (11,0)&0,897&1,488&1,243&(9,9)&1,419&2,009&1,761\\
\hline (12,0)&1,015&1,604&1,356&(10,10)&1,620&2,208&1,957\\
\hline (13,0)&1,132&1,720&1,469&(11,11)&1,820&2,407&2,153\\
\hline (14,0)&1,249&1,836&1,582&(12,12)&2,020&2,606&2,348\\
\hline (15,0)&1,365&1,951&1,695&(13,13)&2,219&2,804&2,544\\
\hline (16,0)&1,481&2.066&1.808&(14,14)&2,417&3,001&2,740\\
\hline (17,0)&1,596&2.180&1.921&(15,15)&2,615&3,199&2,936\\
\hline (18,0)&1,711&2.295&2.034&(16,16)&2,813&3,396&3,131\\
\hline (19,0)&1,826&2.409&2.147&(17,17)&3,010&3,593&3,327\\
\hline (20,0)&1,940&2.524&2.260&(18,18)&3,208&3,790&3,523\\
\hline (21,0)&2,055&2.638&2.373&(19,19)&3,405&3,987&3,718\\
\hline (22,0)&2,169&2.752&2.486&(20,20)&3,602&4,184&3,914\\
\hline (23,0)&2,283&2.866&2.599&(21,21)&3,799&4,380&4,110\\
\hline (24,0)&2,397&2.980&2.712&(22,22)&3,995&4,577&4,305\\
\hline (25,0)&2,512&3.093&2.825&(23,23)&4,192&4,773&4,501\\
\hline (26,0)&2,625&3.207&2.938&(24,24)&4,388&4,970&4,697\\
\hline (27,0)&2,739&3.321&3.051&(25,25)&4,585&5,166&4,893\\
\hline (28,0)&2,853&3.435&3.164&(26,26)&4,781&5,362&5,088\\
\hline (29,0)&2,967&3.548&3.277&(27,27)&4,978&5,559&5,284\\
\hline (4,4)&0,364&0,992&0,783&(28,28)&5,174&5,755&5,480\\
\hline (5,5)&0,588&1,200&0,979&(29,29)&5,370&5,951&5,675\\
\hline
\end{tabular}
\end{center}

 * R$^i$$^n$(Cl) and R$^o$$^u$$^t$(Cl) - radii of "inner" and "outer" chlorine
 cylinders.

\end{center}
\end{table}

\begin{table}
\begin{center}
\caption{Indices of intra-atomic bonds (COOPs, e) for some
NiCl$_2$ nanotubes.}
\begin{center}
\begin{tabular}{|c|c|c|c|}
\hline Nanotubes& *Ni-Cl$^i$$^n$& Ni-Cl$^i$$^n$& Ni-Cl$^o$$^u$$^t$\\
\hline (22,0)&0,281&0,242&0\\
\hline (23,0)&0,280&0,244&0\\
\hline (24,0)&0,287&0,244&0\\
\hline (25,0)&0,285&0,248&0\\
\hline (26,0)&0,288&0,248&0\\
\hline (27,0)&0,293&0,247&0\\
\hline (28,0)&0,291&0,250&0\\
\hline (29,0)&0,290&0,253&0\\
\hline (22,22)&0,104&0&0,241\\
\hline (23,23)&0,105&0&0,243\\
\hline (24,24)&0,105&0&0,243\\
\hline (25,25)&0,105&0&0,241\\
\hline (26,26)&0,105&0&0,241\\
\hline (27,27)&0,106&0&0,242\\
\hline (28,28)&0,106&0&0,240\\
\hline (29,29)&0,107&0&0,239\\
\hline
\end{tabular}
\end{center}
*Ni-Cl$^i$$^n$ and Ni-Cl$^o$$^u$$^t$ - couplings of Ni with atoms
of "inner" and "outer" chlorine cylinders. For Ni-Cl$^i$$^n$
bonds, COOPs values for non-equivalent bonds "along" and "across"
the tube axis are given.
\end{center}
\end{table}

\begin{figure}
\includegraphics[width=0.65\textwidth]{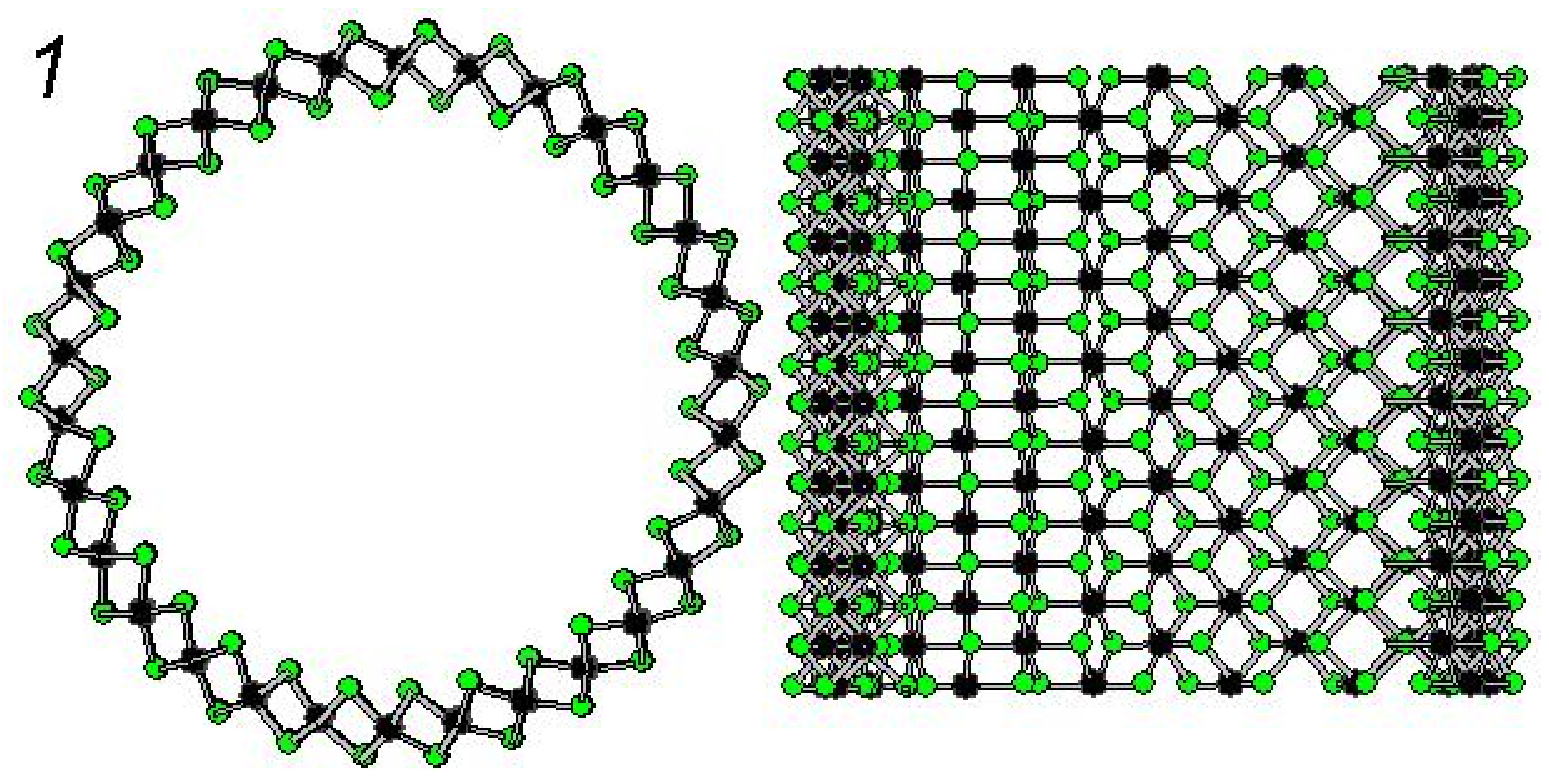}
\includegraphics[width=0.65\textwidth]{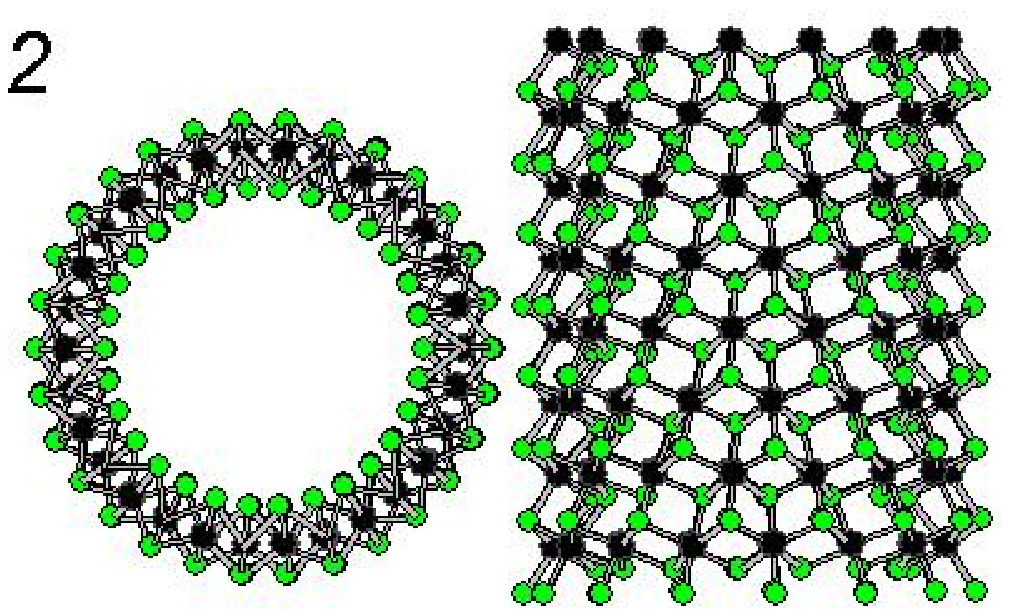}
\caption{Atomic structures of: {\it 1} - {\it armchair} (15,15)-
and {\it 2} - {\it zigzag} (15,0)-like NiCl$_2$ NTs. Side views
and views along the tube axis are shown.}
\end{figure}

\begin{figure}
\includegraphics[width=0.65\textwidth]{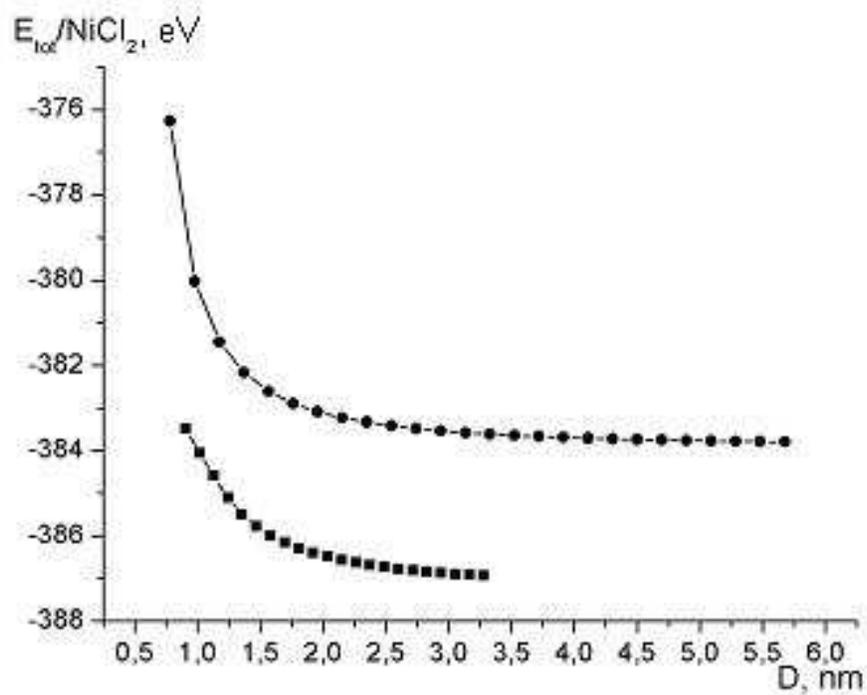}
\caption{Total energies (per NiCl$_2$ unit) as a function of the
diameter of Ni cylinders for NiCl$_2$ tubes of {\it armchair}
({\it n},{\it n}) ({\it rings}) and {\it zigzag} ({\it n},0)
({\it squares}) forms.}
\end{figure}

\begin{figure}
\includegraphics[width=0.65\textwidth]{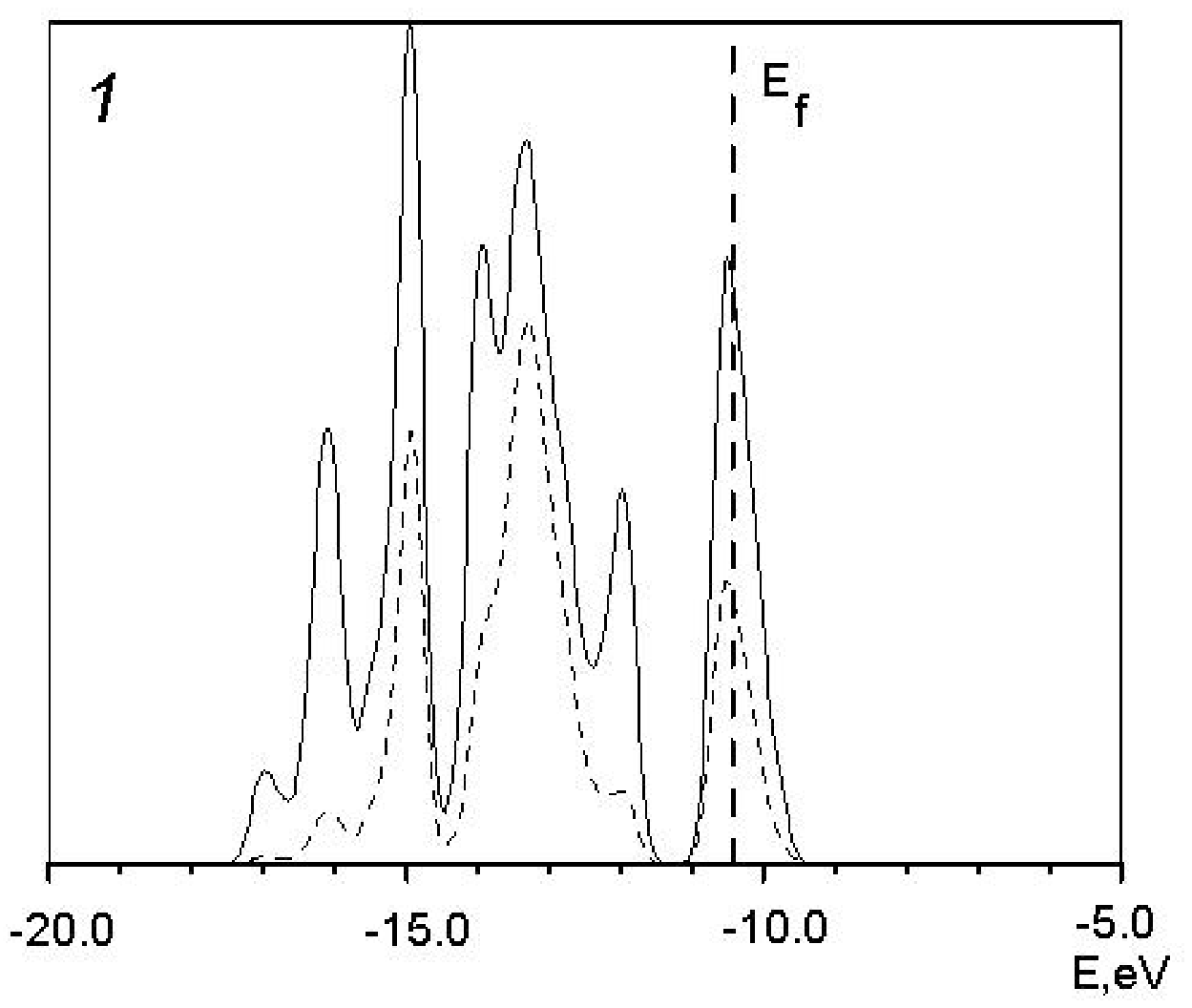}
\includegraphics[width=0.65\textwidth]{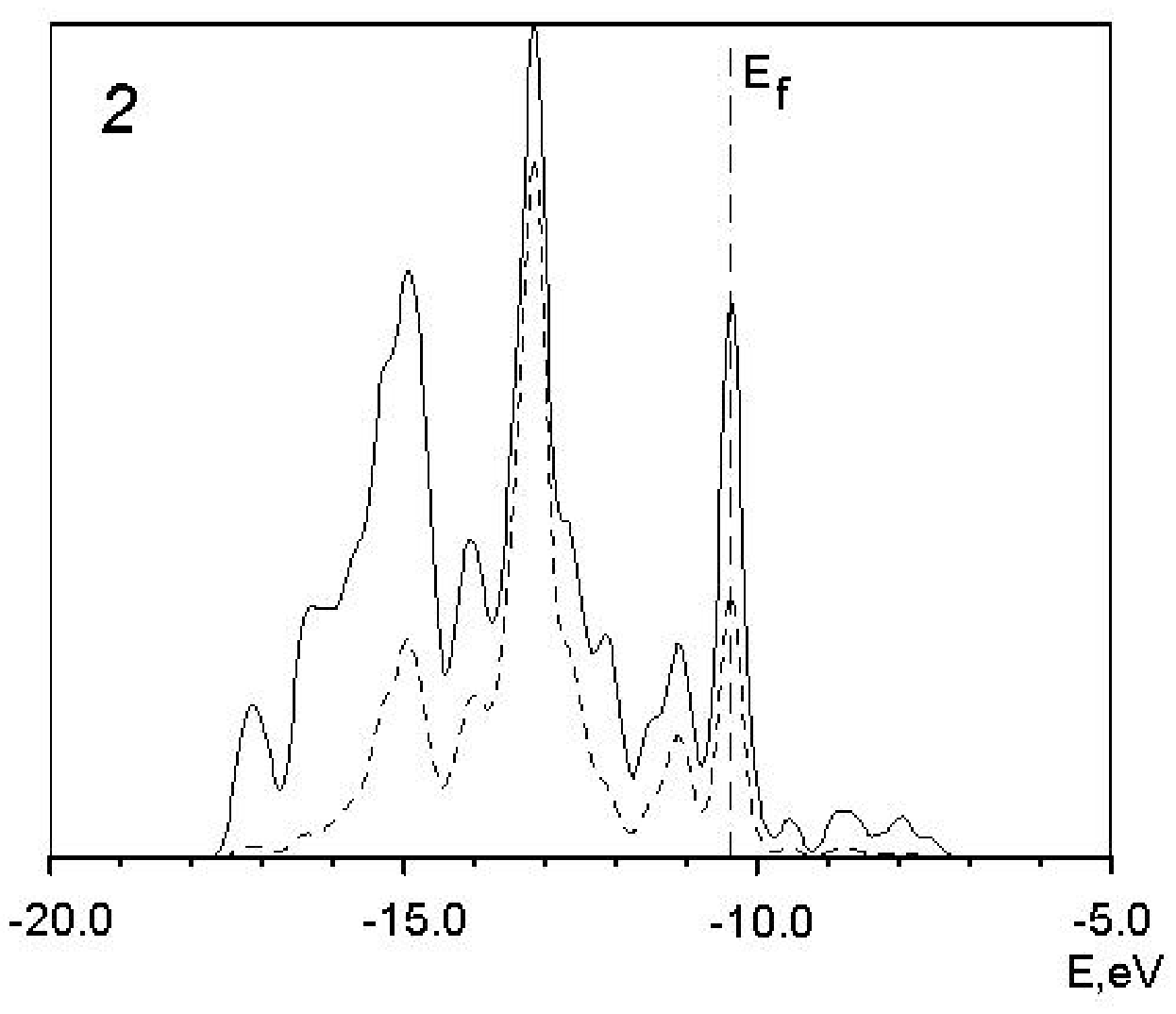}
\caption{Total ({\it solid line}) and Ni3d-like DOS ({\it dotted
line}) of: {\it 1} - {\it armchair} (15,15)- and {\it 2} - {\it
zigzag} (15,0)-like NiCl$_2$ nanotubes.}
\end{figure}

\end{document}